\documentclass[letterpaper, 10 pt, conference]{ieeeconf}  

\IEEEoverridecommandlockouts                              
\usepackage{amsmath} 
\usepackage{amssymb}  
\usepackage{graphicx}

\title{\LARGE \bf
Sampled-data $H^{\infty}$ Optimization for Self-interference Suppression in\\
Baseband Signal Subspaces
}

\author{Hampei Sasahara, Masaaki Nagahara, Kazunori Hayashi, and Yutaka Yamamoto%
\thanks{This research is supported in part by the JSPS Grantin-
Aid for Scientific Research (B) No. 24360163 and (C)
No. 24560543, Grant-in-Aid for Scientific Research on Innovative
Areas No. 26120521, and an Okawa Foundation Research
Grant}
\thanks{H. Sasahara, M. Nagahara, K. Hayashi, and Y. Yamamoto are with Graduate School of Informatics, Kyoto University, Japan; e-mail: sasahara.h@acs.i.kyoto-u.ac.jp, nagahara@i.kyoto-u.ac.jp, kazunori@i.kyoto-u.ac.jp, yy@i.kyoto-u.ac.jp.}
}

\usepackage{color}
\newtheorem{problem}{Problem}
\newtheorem{theorem}{Theorem}

\begin{document}

\maketitle
\thispagestyle{empty}
\pagestyle{empty}

\begin{abstract}
In this article, we propose a design method of self-interference cancelers 
for wireless relay stations
taking account of the baseband signal subspace.
The problem is first formulated as a
sampled-data $H^\infty$ control problem with a generalized sampler
and a generalized hold, which can be reduced to a discrete-time $\ell^2$-induced 
norm minimization problem.
Taking account of the implementation of the generalized sampler and hold,
we adopt the filter-sampler structure for the generalized sampler,
and the uspampler-filter-hold structure for the generalized hold.
Under these implementation constraints,
we reformulate the problem as a standard discrete-time $H^\infty$
control problem by using the discrete-time lifting technique.
A simulation result is shown to illustrate the effectiveness of the proposed method.
\end{abstract}

\section{INTRODUCTION}
\label{sec:intro}
In wireless communications,
relay stations have been used to relay radio signals
between radio stations that cannot directly communicate
with each other due to the signal attenuation.
More recently, relay stations are used to achieve spatial diversity
called cooperative diversity to cope with fading channels \cite{Laneman}.
On the other hand,
it is important to efficiently utilize the scarce bandwidth
due to the limitation of frequency resources
\cite{Cov}, while conventional relay stations commonly use
different wireless resources, such as frequency, time and code,
for their reception and transmission of the signals.
For this reason, a single-frequency full-duplex relay station,
in which signals with the same carrier frequency are received
and transmitted simultaneously, is considered as one of key technologies
in the fifth generation (5G) mobile communications systems \cite{2020beyond}.
In order to realize such full-duplex relay stations,
{\em self-interference} caused by coupling waves is the key issue \cite{Jain}.

\begin{figure}[t]
		\centering
		\includegraphics[width = \linewidth]{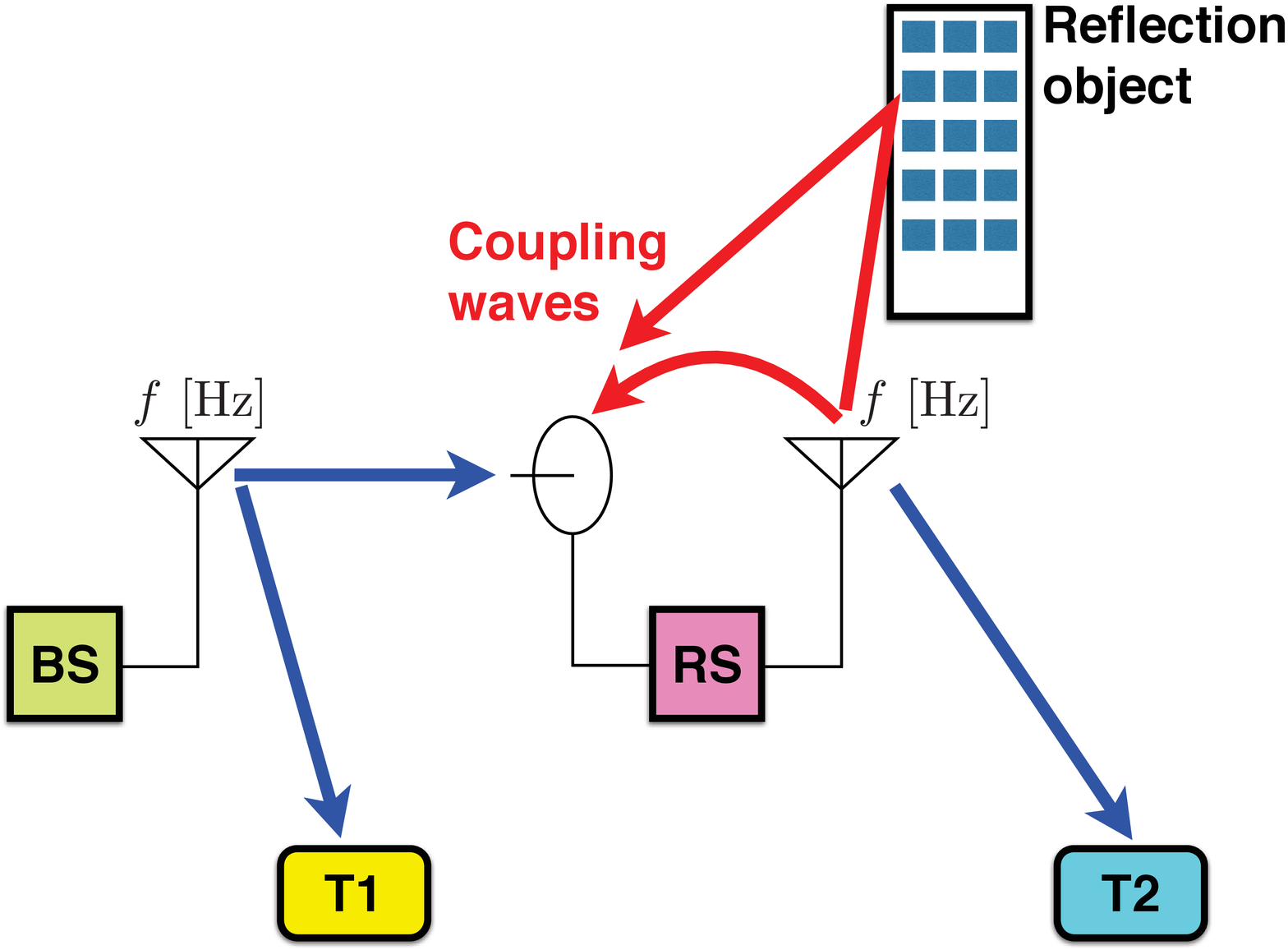}
		\caption{Self-interference}
		\label{coupling}
	\end{figure}

Fig.~\ref{coupling} illustrates self-interference by coupling waves.
In this figure, radio signals with carrier frequency $f$~[Hz]
are transmitted from the base station (denoted by BS).
One terminal (denoted by T1) directly receives the signal from the base station,
but the other terminal (denoted by T2) 
is so far from the base station that they cannot communicate directly.
Therefore, a relay station (denoted by RS) is attached between them
to relay radio signals.
Then, radio signals generated by BS with the same carrier frequency $f$~[Hz]
are transmitted from RS to T2, but also they
are fed back to the receiving antenna of RS
directly or through reflection objects.
As a result, self-interference is caused in the relay station,
which may deteriorate the quality of communication
and, even worse, may destabilize the closed-loop system.

To solve this problem, the authors have proposed a design method of self-interference cancelers that
stabilize the closed-loop system and also suppress the continuous-time effects of self-interference,
based on sampled-data $H^\infty$ control~\cite{SSHR14_01}.
This method takes transmitted signals as filtered $L^2$ signals and minimizes the $H^\infty$ norm
(or the $L^2$-induced norm) of the error system between received and transmitted signals.
However, in many cases, signals used in communications
belong to a subspace of $L^2$ characterized by basis functions.
This subspace is much smaller than the space of filtered $L^2$ signals and hence the method
becomes conservative.

In this paper, we assume that we know the signal subspace
spanned by known basis functions, and utilize the characteristic
of the subspace for self-interference canceler design.
The problem of the canceler design is formulated by
sampled-data $H^\infty$ control problem with generalized sampler/hold.
We show that the problem is reducible to a standard discrete-time $H^\infty$ control problem,
which can be easily solved by numerical computation.
We also show simulation results to illustrate the effectiveness of the proposed method.

The remainder of this paper is organized as follows.
In Section \ref{sec:sm}, we derive a mathematical model of a relay station considered in this study.
In Section \ref{sec:fd}, we formulate the design problem as a sampled-data $H^\infty$ control problem,
which is transformed to a standard discrete-time $H^{\infty}$ control problem.
In Section \ref{sec:ex}, an example is shown to illustrate the effectiveness of the proposed method.
In Section \ref{sec:conc}, we offer concluding remarks.

\subsection*{Notation}
Throughout this paper, we use the following notation.
We denote by $L^2$ the Lebesgue space consisting of all square integrable functions on $[0,\infty)$ 
endowed with the inner product 
\[
 \langle x, y \rangle \triangleq \int_0^\infty y(t)^\top x(t)dt,
\]
and the
norm $\| x \|\triangleq \sqrt{\langle x,x \rangle}$.
We also denote by $\ell^2$ the set of all square summable sequences on $\{0,1,2,\ldots\}$
endowed with the inner product 
\[
\langle x, y \rangle_{\ell^2} \triangleq \sum_{k=0}^\infty y[k]^\top x[x]
\]
and the norm $\| x \|_{\ell^2}\triangleq \sqrt{\langle x,x \rangle_{\ell^2}}$.
The symbol $t$ denotes the argument of time,
$s$ the argument of Laplace transform,
and $z$ the argument of $Z$ transform.
These symbols are used to indicate whether a signal or a system is of continuous-time or discrete-time. 
The operator $e^{-Ls}$ with nonnegative real number $L$ denotes a continuous-time delay operator
and the operator $z^{-l}$ with nonnegative integer $l$ denotes a discrete-time shift operator.


A transfer function of a discrete-time system with state-space matrices $A,B,C,D$ is denoted by
\[
	\left[
	\begin{array}{c|c}
	A & B \\ \hline
	C & D \\
	\end{array}\right] =
	C(zI-A)^{-1}B+D.
\]

\section{SYSTEM MODEL}
\label{sec:sm}

In this section, we derive a mathematical model of a relay station with coupling waves.

There are mainly two types of relaying protocols in wireless communications \cite{Nabar04}.
One protocol demodulates received signals into digital signals,
which are then amplified and transmitted.
This is called an amplify-and-forward (AF) protocol.
The other protocol is called a decode-and-forward (DF) protocol, in which the received signals are demodulated, decoded and detected before transmission.
The most essential difference between the AF protocol and the DF protocol is whether to detect 
received signals or not.
In the DF protocol, 
the detection may reduce the likelihood of instability.
However, the AF protocol requires quite lower implementation complexity than the DF protocol \cite{Nabar04}.
Throughout this study we assume that relay stations operate with the AF protocol.

\begin{figure}[t]
\centering
\includegraphics[width = \linewidth]{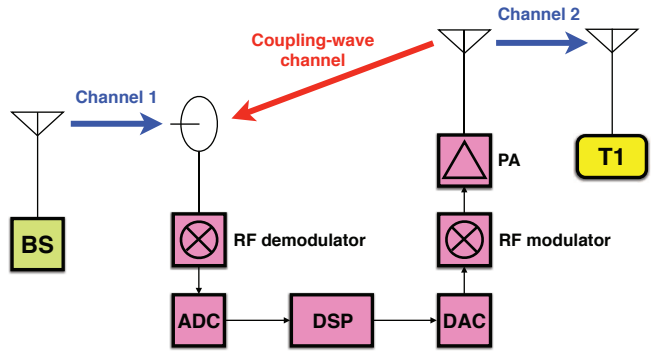}
\caption{AF single-frequency full-duplex wireless relay system}
\label{relay}
\end{figure}

Fig.~\ref{relay} depicts a typical AF single-frequency full-duplex relay station implemented with a digital canceler.
A radio wave from a base station (BS) is accepted at the receiving antenna through Channel 1.
Then, the radio frequency (RF) signal is demodulated to a baseband signal by the RF demodulator,
and converted to a digital signal by the analog-to-digital converter (ADC).
The obtained digital signal is then processed and the effect of loop-back interference mentioned in the previous section is suppressed by the digital signal processor (DSP),
which is converted to an analog signal by the digital-to-analog converter (DAC).
Finally, the analog signal is modulated to an RF signal,
amplified by a power amplifier (PA), and transmitted to terminals (T) with which we communicate by the transmission antenna through Channel 2.
Inevitably, this signal is also fed back to the receiving antenna of the relay station.
This is called coupling wave and causes loop-back interference,
which deteriorates the communication quality.
Practically, 
radio waves may vary because of noise and fading on Channel 1 and Channel 2.
In this study, to concentrate the effect of suppressing coupling
waves, we assume that the characteristics of the channel 1 and 2 in Fig.~\ref{relay}, are modeled as the identity.
It is notable that the assumption is not restrictive when the characteristic of the channels are
known (e.g. channels affected by frequency-selective fading). In this case,
we can include the channel model into the system model.

\begin{figure}[t]
\centering
\includegraphics[width = .8\linewidth]{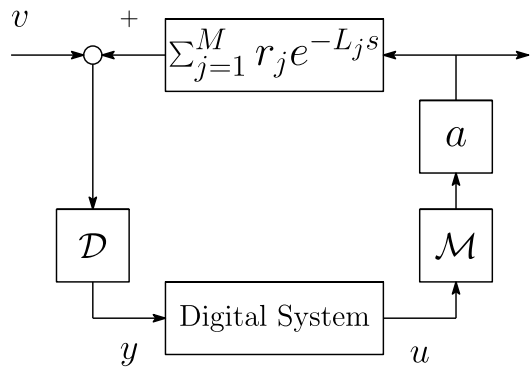}
\caption{Mathematical model of the wireless relay system}
\label{relay_math}
\end{figure}

Fig.~\ref{relay_math} shows a simplified block diagram of the relay station.
We model PA in Fig.~\ref{relay_math} as a static gain denoted by $a$.
The RF modulator is denoted by
${\mathcal M}$ and the RF demodulator by ${\mathcal D}$.
In this study, we adopt a multipath delay system, $\sum_{j=1}^M r_j e^{-L_j s}$, as a channel model,
where  $M$ is the multipath number, $r_j>0$ is the gain, and $L_j>0$ is the delay time of the $j$-th path.
The block named ``Digital System'' includes ADC, DSP and DAC in Fig.~\ref{relay}.

In this article, we consider the quadrature amplitude modulation (QAM), 
which is used widely in digital communication systems, as a baseband modulation method.
RF modulation of QAM transforms a transmission symbol into two orthogonal carrier waves,
that is, a sine wave and a cosine wave.
We assume that the transmission signal $u(t)$, 
which is the output of ``Digital System'' in Fig.~\ref{relay_math},
is given by
\[
	u(t) = \sum_{n=0}^\infty
	 u_d[n]\phi(t-nT),~
	 u_d[n] \triangleq \begin{bmatrix}u_1[n]\\u_2[n]\end{bmatrix}\in{\mathbb R}^2,
\]
where $\phi\in L^2$ is a basis function
(also known as a pulse-shaping function),
and $T$ is the symbol period,
The elements $u_1[n]$, $u_2[n]$
are respectively the in-phase
and the quadrature-phase components of the $n$-th transmission symbol.
The same assumption is also applied to the received baseband signal from BS
\[
 w \triangleq {\mathcal D}v.
\] 
In this study, we define
\[
 \phi_n(t) \triangleq \phi(t-nT),\quad n=0,1,2,\ldots
\] 
and the set $\{\phi_0,\phi_1,\ldots\}$ is an orthonormal system
in $L^2$, that is,
\[
 \langle \phi_i, \phi_j \rangle = \begin{cases}1, & \text{~if~} i=j,\\ 0, & \text{~otherwise.} \end{cases}
\] 
Note that this assumption is satisfied for a square root raised cosine (SRRC) pulse,
which is widely used in digital communications \cite{Jones01}.

\begin{figure}[t]
\centering
\includegraphics[width = .9\linewidth]{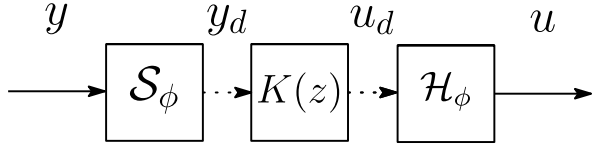}
\caption{Digital system}
\label{DSP}
\end{figure}

Next we model the ``Digital System'' block in Fig.~\ref{relay_math} as a sampler-filter-hold system
shown in Fig.~\ref{DSP}.
In this block diagram,
${\mathcal S}_\phi$ is a generalized sampler defined by
\[
 {\mathcal S}_{\phi}: \left\{\begin{bmatrix}y_1(t)\\y_2(t)\end{bmatrix}\right\}_{t\in[0,\infty)} \mapsto 
 \left\{\begin{bmatrix}\langle y_1, \phi_n \rangle\\ \langle y_2, \phi_n\rangle\end{bmatrix}\right\}_{n=0,1,2,\ldots}.
\]
$K(z)$ is a digital filter that we design for self-interference cancelation.
${\mathcal H}_{\phi}$ is a generalized hold defined by
\begin{gather*}
	{\mathcal H}_{\phi}: \left\{\begin{bmatrix}u_1[n]\\u_2[n]\end{bmatrix}\right\}_{n=0,1,2,\ldots} \mapsto
	\left\{\sum_{n=0}^{\infty} \begin{bmatrix}u_1[n]\\u_2[n]\end{bmatrix} \phi_n(t)\right\}_{t\in[0,\infty)}.
\end{gather*}

\begin{figure}[t]
\centering
\includegraphics[width = .9\linewidth]{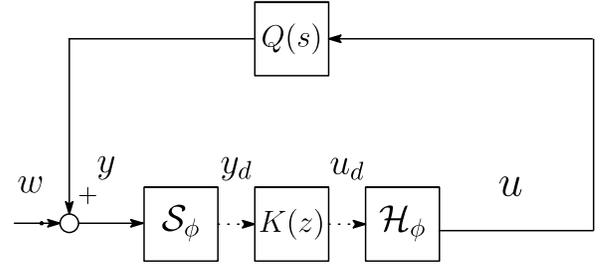}
\caption{Transformed wireless relay system}
\label{relay_system}
\end{figure}

Finally, after some transformations \cite{SSHR14_01}, 
we obtain a relay station model depicted in Fig.~\ref{relay_system},
which is equivalent to Fig.~\ref{relay_math}.
In this block diagram, we define
\[
 Q(s) \triangleq \sum_{j=1}^M ar_j e^{-L_js}
  \begin{bmatrix}
	\cos 2\pi fL_j & \sin 2\pi fL_j \\
	-\sin 2\pi fL_j & \cos 2\pi fL_j \\
  \end{bmatrix}.
\]	
The input signal $w$ is defined by $w={\mathcal D}v$ as mentioned above,
and we assume that $w$ can be represented by a linear combination of $\phi_0,\phi_1,\ldots$.
By the block diagram, we can see that the relay station with self-interference is a \emph{feedback} system.
In practice, the gain of the power amplifier (PA) in Fig.~\ref{relay_system} is very high
(e.g.,  $a=1000$) and the loop gain becomes much larger than $1$,
and hence we should discuss the \emph{stability} as well as self-interference cancelation.
To achieve these requirements, 
we formulate the design problem as a sample-data $H^\infty$ control problem.

\section{PROBLEM FORMULATION}
\label{sec:fd}

In this section, we formulate the design problem as a sampled-data $H^\infty$ control problem
based on the model discussed in the previous section.
Then we show that the problem is reduced to a problem of
discrete-time $\ell^2$-induced norm minimization.

\subsection{Design Problem via Sampled-Data $H^{\infty}$ control}
\begin{figure}[t]
\centering
\includegraphics[width = .8\linewidth]{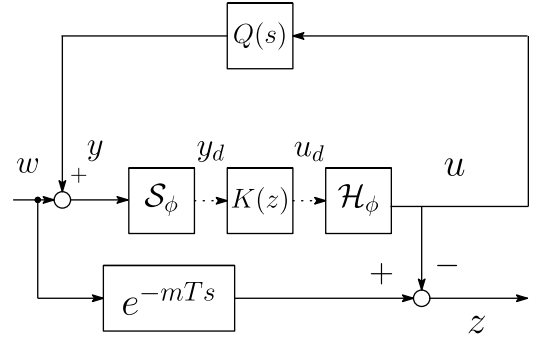}
\caption{Sampled-data error system}
\label{errorSD}
\end{figure}

The objective is to reduce the error between the input $w$ and the output $u$ shown 
in Fig.~\ref{relay_system}.
In many cases, the output $u$ is allowed to be delayed against the input $w$,
and hence we consider the delayed error
\[
 z(t) \triangleq w(t-mT)-u(t),
\]
where $m$ is a non-negative integer.
The corresponding error system is shown in Fig.~\ref{errorSD}. 

Then, as mentioned above, the received signal $w$
is represented by a linear combination of $\phi_n$.
Let us define the signal subspace $W$ by
\begin{equation}
 W \triangleq \overline{\rm{span} \{\phi_0,\phi_1,\ldots\}}
\end{equation}
where span$\{\phi_0,\phi_1,\ldots\}$ is the space consisting of
all finite linear combinations of $\phi_0$, $\phi_1$,\ldots,
and the overline denotes the closure in $L^2$.

Now, we formulate our design problem.
\begin{problem}
\label{Problem1}
Find the digital controller (canceler) $K(z)$ that stabilizes the feedback system in Fig.~\ref{errorSD}
and minimizes the following cost function:
\[
 J \triangleq \sup_{w\in W, \|w\|=1} \|z\|.
\]
\end{problem}
This is a sampled-data $H^\infty$ control problem.
In the next subsection, we reduce this problem as
a problem of discrete-time $\ell^2$-induced norm minimization.


\subsection{Transform into Discrete-Time $H^{\infty}$ Control Problem}
\label{subsec:pdp}
As discussed above, the received signal $w$ is in the subspace $W$.
Since the delay $mT$ is an integer multiple of $T$,
we can represent the delayed input $w(t-mT)$ as
\begin{equation}
	w(t-mT) = \sum_{n=0}^{\infty} w_d[n] \phi_n(t),
\label{def:w_d}
\end{equation}
where $w_d[n]\in{\mathbb R}^2$ is the $n$-th coefficient.
Also, since the transmitted signal $u$ is the output of ${\mathcal H}_\phi$,
$u$ is also in $W$. Thus, we have
\[
	u(t) = \sum_{n=0}^{\infty} u_d[n] \phi_n(t),
\]
where $u_d[n]\in{\mathbb R}^2$ is the $n$-th coefficient.
Define
\begin{eqnarray}
	z_d[n] \triangleq w_d[n] - u_d[n], \quad n=0,1,2,\ldots
\label{def:z_d}
\end{eqnarray}
Then we have
\[
	z(t) = w(t-mT)-u(t) = \sum_{n=0}^{\infty} z_d[n] \phi_n(t).
\]

\begin{figure}[t]
\centering
\includegraphics[width = \linewidth]{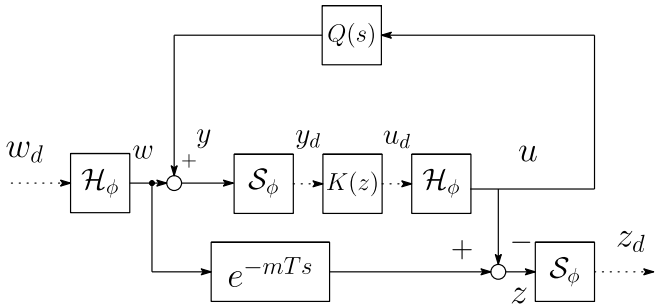}
\caption{Discrete-time error system}
\label{errorD}
\end{figure}

Now we have the following theorem:
\begin{theorem}
The sampled-data error system shown in Fig.~\ref{errorSD}
is equivalent to the error system shown in Fig.~\ref{errorD}
in the sense that
\[
 J = \sup_{w\in W, \|w\|=1} \|z\| = \sup_{w_d\in \ell^2, \|w_d\|_{\ell^2}=1} \|z_d\|_{\ell^2}.
\]
\end{theorem}
\begin{proof}
Since $W\subset L^2$ is a closed subspace,
$W$ is a Hilbert space.
Since $\rm{span} \{\phi_0,\phi_1,\ldots\}$ is dense in $W$, 
$\{\phi_0,\phi_1,\ldots\}$ is a complete orthonormal system in $W$.
Thus, for any $w\in W$, Perseval's theorem \cite{YYtext} gives
\[
	\|w\| = \|w_d\|_{\ell^2},
\]
where $w_d[n]=\langle w, \phi_n \rangle$.
Also, since $z\in W$ for any $w\in W$, we have
\[
	\|z\| = \|z_d\|_{\ell^2},
\]
where $z_d[n]=\langle z, \phi_n \rangle$.
Then, $w \in W$ is a necessary and sufficient condition for $w_d \in \ell^2$, and hence we have
\[
	\sup_{w\in W, \|w\|=1}\|z\| = \sup_{w_d\in \ell^2, \|w_d\|_{\ell^2}=1} \|z_d\|_{\ell^2}.
\]
\end{proof}

From this theorem, Problem 1 is transformed into
the following discrete-time problem:
\begin{problem}
Find the digital controller (canceler) $K(z)$ that stabilizes the feedback system in Fig.~\ref{errorD}
and minimizes
\[
	J_d \triangleq \sup_{w_d\in \ell^2, \|w_d\|_{\ell^2}=1} \|z_d\|_{\ell^2}.
\]
\label{Problem2}
\end{problem}\

\section{Implementation-Aware Design}
In this section, we will give a design formula for the $H^\infty$-optimal
digital canceler $K(z)$.
At the same time, we take account of implementation of the generalized sampler ${\mathcal S}_\phi$ and
the generalized hold ${\mathcal H}_\phi$.

\begin{figure}[t]
\centering
\includegraphics[width = \linewidth]{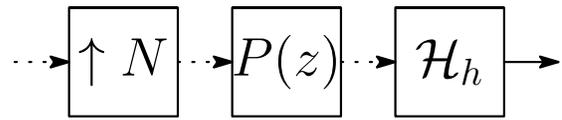}
\caption{The band-limiting and pulse-shaping system $\tilde{{\mathcal H}_h}$}
\label{tildeH}
\end{figure}

\begin{figure}[t]
\centering
\includegraphics[width = \linewidth]{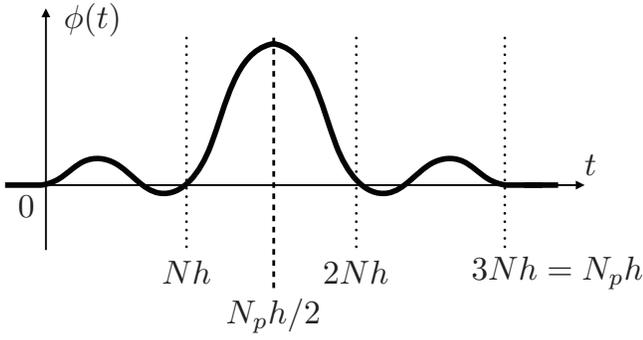}
\caption{Basis function $\phi$ with finite support}
\label{fig:phi}
\end{figure}

\begin{figure}[t]
\centering
\includegraphics[width = \linewidth]{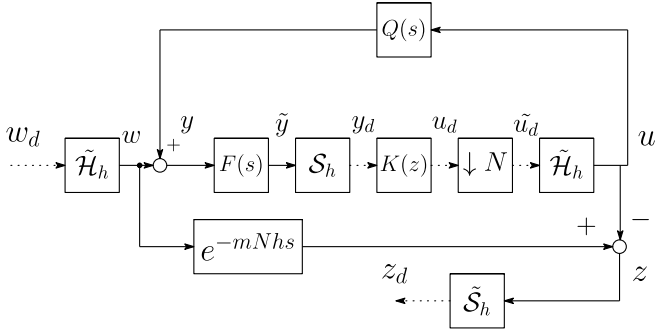}
\caption{Discrete-time error system with $\tilde{{\mathcal H}}_h$}
\label{errorD2}
\end{figure}

In practice, the ideal uniform sampler with an anti-aliasing filter
is widely used in digital systems
in place of the generalized sampler ${\mathcal S}_\phi$.
The ideal (2-dimensional) uniform sampler ${\mathcal S}_h$ with sampling time $h>0$ is defined by
\[
 {\mathcal S}_h: \left\{\begin{bmatrix}y_1(t)\\y_2(t)\end{bmatrix}\right\}_{t\in[0,\infty)} \mapsto 
 \left\{\begin{bmatrix} y_1(nh) \\  y_2(nh) \end{bmatrix}\right\}_{n=0,1,2,\ldots}.
\]
We assume that the sampling time $h$ is determined by
$T=Nh$ for some natural number $N$, that is,
$h=T/N$.
We denote the anti-aliasing filter by $F(s)$, which is assumed to be a 2-input/2-output,
proper, stable transfer function matrix.
That is, the generalized sampler ${\mathcal S}_\phi$ is implemented as ${\mathcal S}_h F(s)$.
On the other hand, the generalized hold ${\mathcal H}_\phi$ is also implemented by
using an upsampler $\uparrow N$ with upsampling ratio $N$,
an FIR (finite impulse response) digital filter $P(z)$, and a zero-order hold ${\mathcal H}_h$
with sampling time $h$, as shown in Fig.~\ref{tildeH}.
Here we define the upsampler $\uparrow N$ by
\begin{equation*}
\uparrow N: \{x[k]\}_{k=0}^{\infty} \mapsto \{x[0],\underbrace{0,\ldots,0,}_{N-1} x[1], 0, \ldots \},
\end{equation*}
and the zero-order hold ${\mathcal H}_h$ by
\begin{gather*}
	{\mathcal H}_h: \left\{\begin{bmatrix}u_{d1}[n]\\u_{d2}[n]\end{bmatrix}\right\}_{n=0,1,2,\ldots} \mapsto
	\left\{\begin{bmatrix}u_1(t)\\u_2(t)\end{bmatrix}\right\}_{t\in[0,\infty)}.
\end{gather*}
We denote the system shown in Fig.~\ref{tildeH} by $\tilde{{\mathcal H}}_h$,
that is,
\begin{equation*}
	\tilde{{\mathcal H}}_{h} \triangleq {\mathcal H}_h P(z) (\uparrow N).
\end{equation*}
This structure is widely used for pulse-shaping \cite{AnalogDevices},
and the filter $P(z)$ is called a band-limiting filter.
The filter $P(z)$ is usually designed such that its impulse response is equal to
the sampled values of the baseband pulse $\phi$.
More precisely, we set
\[
 P(z) = \sum_n \phi(nh)z^{-i}.
\] 
We assume that the support of $\phi(t)$ is finite and the support length is
an integer multiple of $h$, that is, there exists a natural number $N_p$ such that
the support length of $\phi(t)$ is $N_ph$.
See an illustration of $\phi(t)$ shown in Fig.~\ref{fig:phi}.
By this assumption, the filter $P(z)$ becomes an FIR filter.
Fig.~\ref{errorD2} shows the system in which the generalized sampler ${\mathcal S}_\phi$ and the
generalized hold ${\mathcal H}_\phi$ are replaced by ${\mathcal S}_hF(s)$ and $\tilde{{\mathcal H}}_h$,
respectively.

\begin{figure}[t]
\centering
\includegraphics[width = .65\linewidth]{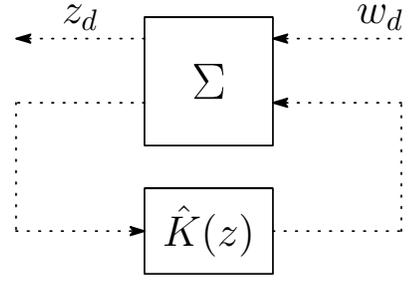}
\caption{The generalized system}
\label{GP}
\end{figure}

Then, let us consider the generalized sampler ${\mathcal S}_\phi$ for the
error $z$ (see Fig.~\ref{errorD}).
This sampler is virtual (i.e., not implemented), and we can adopt
more precise approximation than ${\mathcal S}_hF(s)$.
We here assume that the pulse $\phi(t)$ with support length $N_ph$
has its peak at $t=N_ph/2$, the center of the support set
(see Fig.~\ref{fig:phi}).
Then, we can well approximate ${\mathcal S}_\phi z$ by
$z(N_ph/2)$.
That is, we approximate ${\mathcal S}_\phi$ by $\tilde{{\mathcal S}}_h \triangleq {\mathcal S}_{Nh}e^{(N_ph/2)s}$.
We show the corresponding block diagram in Fig.~\ref{errorD2}
where $\downarrow N$ is a downsampler defined as
\[
	\downarrow N : \{ x[k]\} \mapsto \{ x[0],x[N],x[2N],\ldots \}.
\]
This system has two sampling rates and becomes a multi-rate system,
which can be equivalently transformed into a single-rate system by 
employing the technique called as discrete-time lifting \cite{Chen95}.
The discrete-time lifting of a discrete-time system is defined by
\[
\begin{split}
&\mathbf{lift}\left( \left[
	\begin{array}{c|c}
	A & B \\ \hline
	C & D \\
	\end{array} \right], N \right) \\
	&~\triangleq \mathbf{L}_N \left[
	\begin{array}{c|c}
	A & B \\ \hline
	C & D \\
	\end{array} \right], \mathbf{L}_N^{-1} \\
	&~= \left[ \begin{array}{c|cccc}
	A^N & A^{N-1}B & A^{N-2}B & \ldots & B \\ \hline
	C & D & 0 & \ldots & 0 \\
	CA & CB & D & \ddots & \vdots \\
	\vdots & \vdots & \vdots & \ddots & 0 \\
	CA^{N-1} & CA^{N-2}B & CA^{N-3}B & \ldots & D \\
	\end{array} \right], 
 \end{split}
\]
where
\[
 \begin{split}
	\mathbf{L}_N &\triangleq ( \downarrow N ) [
	\begin{array}{cccc}
	1 & z & \cdots & z^{N-1} \\
	\end{array}]^T, \\
	\mathbf{L}_N^{-1} &\triangleq [
	\begin{array}{cccc}
	1 & z^{-1} & \cdots & z^{-N+1} \\
	\end{array} (\uparrow N).
 \end{split}
\]
Then we have the generalized plant representation shown in Fig.~\ref{GP},
where we define
\[
 \begin{split}
	\Sigma &\triangleq \left[
	\begin{array}{cc}
	T_{11} & T_{12} \\
	T_{21} & T_{22} \\
	\end{array} \right],\\
	\hat{K}(z) &\triangleq \mathbf{L}_N K(z) \mathbf{L}^{-1}_N,\\
	T_{11} &\triangleq {\mathcal S}_{Nh} e^{(N_p/2-mN)hs} {\mathcal H}_h P(z) (\uparrow\! N),\\
	T_{12} &\triangleq -{\mathcal S}_{Nh}e^{N_ph/2s} {\mathcal H}_h P(z) (\uparrow\! N) (\downarrow\! N) \mathbf{L}_N^{-1},\\
	T_{21} &\triangleq \mathbf{L}_N {\mathcal S}_h F(s) {\mathcal H}_h P(z) (\uparrow\! N),\\
	T_{22} &\triangleq \mathbf{L}_N {\mathcal S}_h F(s) Q(s) {\mathcal H}_h
	P(z) (\uparrow\! N) (\downarrow\! N) \mathbf{L}_N^{-1}. \\
 \end{split}
\]
Note that if 
\[
 \frac{N_p}{2}<mN,
\]
then the system $\Sigma$ becomes a linear time-invariant discrete-time system. 
Thus, by using the generalized form, one can obtain the optimal controller $\hat{K}(z)$
to minimize the $H^\infty$ norm of
\begin{equation*}
T_{\mathrm{cl}} \triangleq {\mathcal F}(\Sigma, \hat{K}),
\end{equation*}
via a standard discrete-time $H^\infty$ optimal control method \cite{Chen95}.

\section{Simulation}
\label{sec:ex}
We here show a simulation result of self-interference cancelation 
to illustrate the effectiveness of the proposed method.

We chose the following simulation parameters:
the sampling period $h$ to be normalized to $1$,
the normalized carrier frequency $f$ is $10000$ [Hz],
the symbol period $T=Nh=2$ (the upsampling ratio $N=2$),
and the process delay $m=5$.
We assume the anti-alias analog filter is given by
\[
	F(s) = \dfrac{1}{0.5s+1}I,
\]
and the transmission gain $a = 2500$.
We also assume the multipath number $M$ is $2$,
and its characteristic is $0.2e^{-10s}+0.17e^{-12s}$.
The baseband modulation is taken as the orthogonal frequency division multiplexing with the binary phase shift keying \cite{Goldsmith05},
where we chose the block length as $64$ and the guard length as $16$.

\begin{figure}[t]
\centering
\includegraphics[width = .9\linewidth]{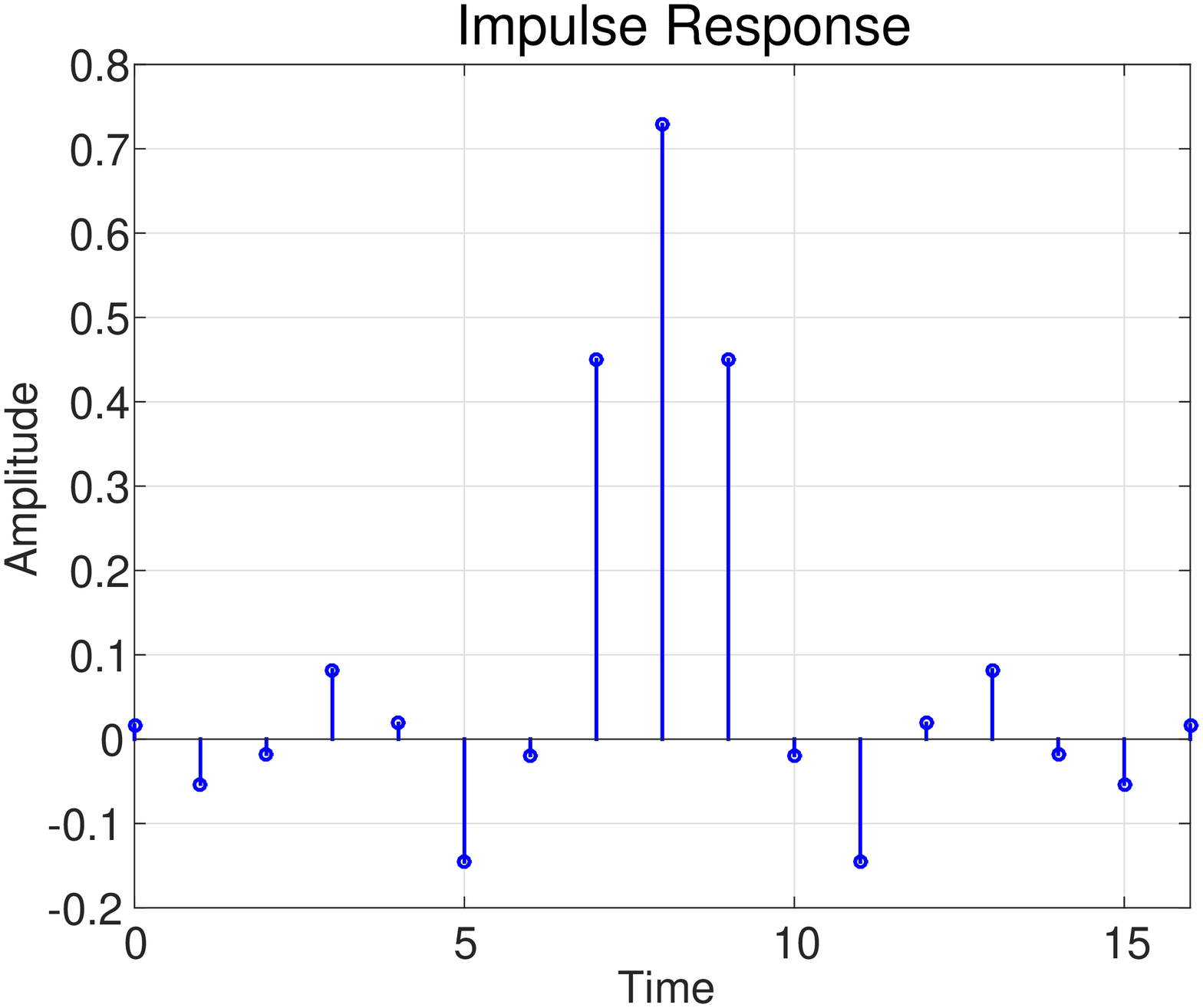}
\caption{Impulse response of the band-limiting filter $P(z)$}
\label{blfilter}
\end{figure}

We employ the band-limiting filter $P(z)$ as
\begin{equation*}
\begin{array}{l}
P(z) = 0.0165 - 0.0533 z^{-1} - 0.0177 z^{-2} + 0.0821 z^{-3} \\
\mbox{}+ 0.0186 z^{-4} - 0.1455 z^{-5} - 0.0192 z^{-6} + 0.4499 z^{-7} \\
\mbox{}+ 0.7287 z^{-8} + 0.4499 z^{-9} - 0.0192 z^{-10} - 0.1455 z^{-11}\\
\mbox{}+ 0.0186 z^{-12} + 0.0821 z^{-13} - 0.0177 z^{-14} - 0.0533 z^{-15} \\
\mbox{}+ 0.0165 z^{-16}. \\
\end{array}
\end{equation*}
This filter is a SRRC band-limiting filter \cite{William03} with roll-off factor $\beta = 0.1$.
Fig.~\ref{blfilter} shows the impulse response of $P(z)$.

\begin{figure}[t]
\centering
\includegraphics[width = .9\linewidth]{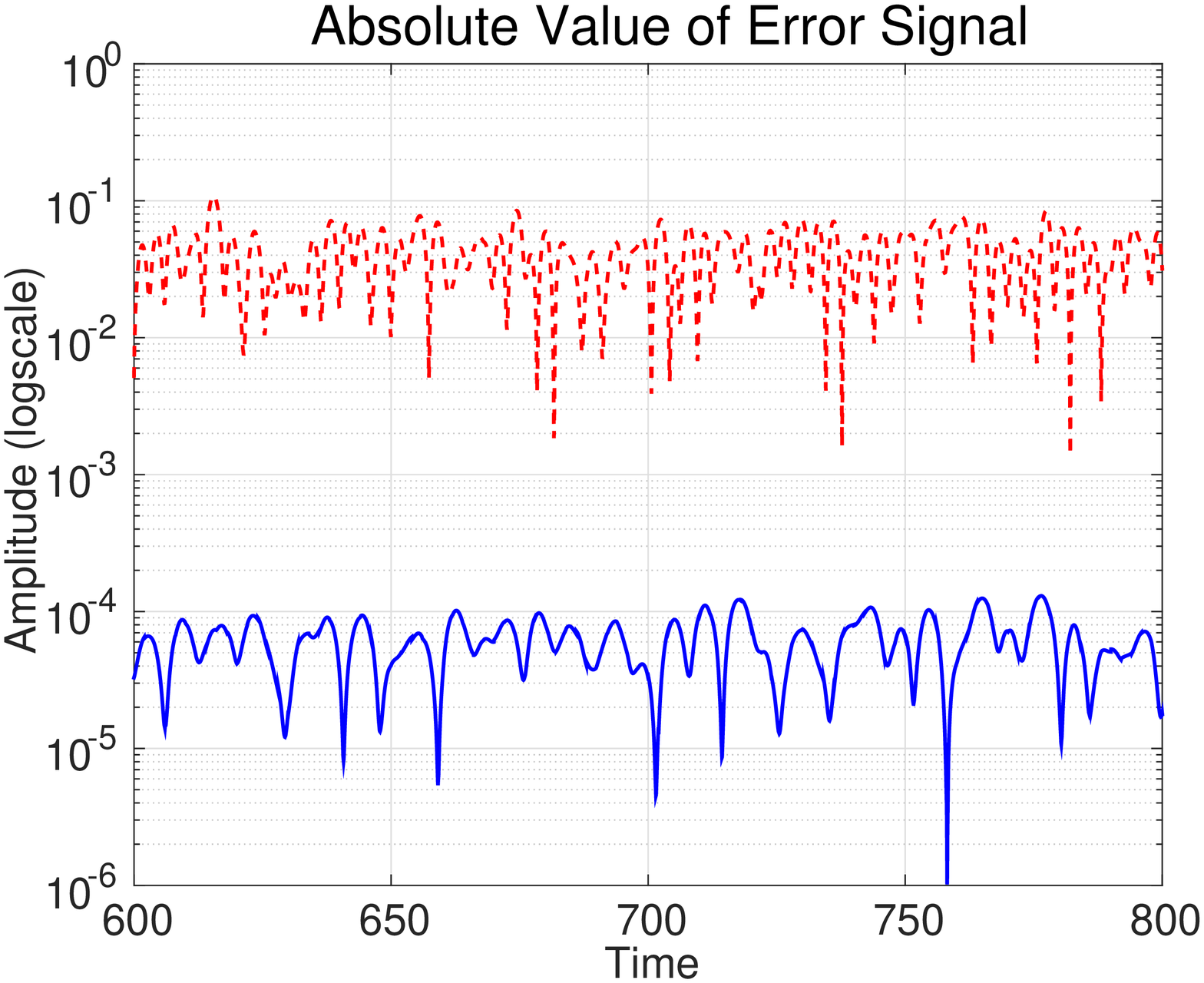}
\caption{Absolute value of error signal $|z(t)|$: with Sampled-Data Method (dash-dot line) and Proposed Method (sold-line)}
\label{sim}
\end{figure}

Fig.~\ref{sim} shows the absolute value of the error signal $z(t)$ with a log-scale vertical axis.
The dash-dot line represents $|z(t)|$ by the canceller designed by a conventional method
proposed in \cite{SSHR14_01},
and the solid line represents $|z(t)|$ by the proposed canceller.
This figure shows that the error is much reduced with the proposed method than with the conventional method.
From this result, our method that takes account of the signal baseband subspace $W$ is
much more effective than the conventional method.

\section{Conclusion}
\label{sec:conc}

In this article, we have proposed a new method for self-interference cancelation 
taking account of the baseband signal subspace.
The problem is first formulated as a
sampled-data $H^\infty$ control problem with a generalized sampler
and a generalized hold, which can be reduced to a discrete-time $\ell^2$-induced 
norm minimization problem.
Taking account of the implementation of the generalized sampler and hold,
we adopt the filter-sampler structure for the generalized sampler,
and the uspampler-filter-hold structure for the generalized hold.
Under these implementation constraints,
we reformulate the problem as a standard discrete-time $H^\infty$
control problem by using the discrete-time lifting technique.
A simulation result is shown to illustrate the effectiveness of the proposed method.

\bibliographystyle{IEEEtran}
\bibliography{sshrrefs}


\end{document}